\documentclass[preprint,12pt]{elsarticle}

\usepackage{hyperref}
\usepackage{amssymb}
\usepackage{amsthm,booktabs,graphicx,adjustbox}
\let\counterwithout\relax

\usepackage{chngcntr}

\journal{Environmental Research}

\begin{document}

\begin{frontmatter}


\title{The short-term seasonal analyses between atmospheric environment and COVID-19 in epidemic areas of Cities in Australia, South Korea, and Italy}
% \tnotetext[label1]{Australia}
% \author{Yuxi Liu\corref{cor1}\fnref{label2}}
% \ead{liu1356@flinders.edu.au}
% \fntext[label2]{Asio}
% % \cortext[cor1]{Sant Crack}
% \address{College of Science and Engineering, Flinders University, Adelaide, Australia\corref{cor1}}
% \fntext[label3]{Dopjsi}
\author[label1]{Yuxi Liu}
\address[label1]{College of Science and Engineering, Flinders University, Adelaide, Australia}
%\ead{liu1356@flinders.edu.au}
\author[label2]{Xin Lin}
\address[label2]{School of Computer Science and Engineering, Beihang University, Beijing, China}
% \ead{sweeneylin@buaa.edu.cn}
\author[label1]{Shaowen Qin\corref{cor1}}
\ead{shaowen.qin@flinders.edu.au}
\cortext[cor1]{Corresponding author}

% \title{The short-term seasonal impact between atmospheric environment and COVID-19 in epidemic areas of Cities in Australia, South Korea, and Italy}


% \author{}

\address{}
% \clearpage

\begin{abstract}
The impact of the outbreak of COVID-19 on health has been widely concerned. Disease risk assessment, prediction, and early warning have become a significant research field. Previous research suggests that there is a relationship between air quality and the disease. This paper investigated the impact of the atmospheric environment on the basic reproduction number (R$_0$) in Australia, South Korea, and Italy by using atmospheric environment data, confirmed case data, and the distributed lag non-linear model (DLNM) model based on Quasi-Poisson regression. The results show that the air temperature and humidity have lag and persistence on short-term R$_0$, and seasonal factors have an apparent decorating effect on R$_0$. PM$_{10}$ is the primary pollutant that affects the excess morbidity rate. Moreover, O$_3$, PM$_{2.5}$, and SO$_2$ as perturbation factors have an apparent cumulative effect. These results present beneficial knowledge for correlation between environment and COVID-19, which guiding prospective analyses of disease data.

\end{abstract}

% Graphical abstract
%   \includegraphics[width=1.\columnwidth]{./figures/spa.pdf}

\maketitle

\end{frontmatter}


\section{Introduction}
% \label{}
At the end of 2019, the outbreak of Coronavirus disease (COVID-19) pandemics caused global panic. COVID-19 epidemic has spread throughout the world due to the long incubation period and highly infectious. By the end of April 17th, 2020, there were a total of 2034802 confirmed cases, 135163 confirmed deaths, and 213 counties, areas, or territories with cases \cite{cite1}. Until reliable vaccines become available, public health responses to COVID-19 are focused on hand hygiene, social distancing, and quarantine strategies to reduce community transmission rates. Based on the report \cite{cite2} issued by the World Health Organization (WHO), the main transmission channels of COVID-19 are droplet transmission and close contact transmission, and aerosol transmission. The general spread of influenza viruses primarily occurs through respiratory droplets between persons within 1 meter. Transmission might therefore be reduced in the warmer and more humid weather when droplets fall to the ground more easily. Similarly, a 15$\%$ increase in air pollutants such as PM$_{2.5}$, PM$_{10}$, SO$_2$, CO, NO$_2$, and O$_3$ can increase the risk of influenza by approximately 1.0$\%$ \cite{cite3}. On May 1st, 2020, Bharat Pankhania, an infectious disease expert at the University of Exeter, UK, said that the vital factor determining epidemic transmission is the environment \cite{cite4}. While the epidemic situation outbreak began in the northern hemisphere, some countries in the southern hemisphere are entering winter due to seasonal changes between northern and southern hemispheres, such as Australia. Furthermore, most of the epidemic regions are located in cities with severe air pollution. Therefore, it is necessary to investigate the relationship between COVID-19 and atmospheric environment.

The traditional approaches to analyze the impact of the atmospheric environment on disease, including Time-Series Analysis and Case Cross-Analysis, in which time series analysis can divide into simple Poisson regression, traditional generalized linear model (GLM), and the generalized additive model (GAM). A significant analysis and discussion on the exposure toxicants were presented by Gordon and Leon \cite{cite9}, which attempts to review the environmental heat stress challenge, and how they affect the thermoregulation and related pathophysiological responses to environmental toxicants by time-Series Analysis model. Peng et al. \cite{cite10} conducted a Bayesian semiparametric hierarchical model that estimates the time-varying effect of air pollution on daily mortality. To further investigate particulate pollutants' impact on cardiovascular disease, Leon \cite{cite11} employed the classical time series analysis approach to modeling. Kleinman et al. \cite{cite16} applying generalized linear mixed models implement a general approach to evaluating whether observed counts in relatively small areas are larger than would be expected based on a history of naturally occurring disease. Brook et al. \cite{cite13} studied the structural behavior of PM$_{2.5}$ exposure impact on cardiovascular incidence rate was carried out by Generalized Additive Models. To quantify the effect of stimulus in the presence of covariate data, Zanobetti \cite{cite14} integrated generalized additive model and distributed lag model and proposed generalized additive distributed lag models (DLNM). Armstrong \cite{cite15} introduced the DLNM model into the study of temperature health effects and simulated a series of relationships between multi lag nonlinear temperature and mortality. Gasparrini et al. \cite{cite12} further develop on the DLNM model, which described the theory of distributed lag nonlinear model using the cross-basis. The basic reproduction number R$_0$ describes the mean number of infections generated during the infectious period of a single infective when there is no immunity from past exposures of vaccination \cite{cite5}. The COVID-19 case occurred in Australia as early as January 26. However, it was not until the end of March that Australia imposed strict control policies. Given that infected and confirmed cases are one to two weeks leading of time, there is a significant reduction in new cases in early April before these policy interventions take effect. Instead, they indicated the environment that existed before policy measures were offered \cite{cite19}, which tends to suggest that Australia had a pre-shutdown R$_0$ well below that seen in most other parts of the world. Australia seems to have a relatively strong natural defense against the virus, but accounting for these different relationships is challenging.

In this paper, we apply the DLNM model for estimating seasonal and air pollution patterns effects on R$_0$ in multisite time-series studies and estimating time-varying COVID-19 effects within each city and for comparing temporal patterns across cities and geographic regions. Using data from the CSSEGISandData \cite{cite6} case data, weather city data from the Air Quality Open Data Platform \cite{cite8}, we estimate the effects of particulate matter 14ug/m$^{3}$ in aerodynamic diameter (PM$_{10}$) on daily R$_0$. The seasonal patterns are estimated for ten cities and on average for Australia, South Korea, and Italy. We explore the sensitivity of estimated seasonal, air pollutants, and exposure lag (in days) patterns for short-term R$_0$ trends.

\section{Materials and methods}

\section*{2.1. Database description}
Eligible atmospheric environment data that matched the selection criteria were collected by Air Quality Open Data Platform, which includes daily time series on weather and air pollution assembled from publicly available sources for the significant largest cities in the global countries. The most recent data are available on the website (https://aqicn.org/data-platform). Data on confirmed cases and deaths for the experiment phase of the project sourced from publicly available sources including the John Hopkins GitHub COVID-19 repository \cite{cite17}, which includes comprehensive data for confirmed cases and deaths at the State level and data are available at the website of the COVID-19 time-series data \cite{cite18} (https://blog.stata.com).

\section*{2.2. Statistical methods}
The data were normalized using confirmed case data of some cities with severe epidemics situation in Australia(Autumn season), South Korea(Spring season), and Italy(Spring season). R$_0$ was estimated using R language package EpiEstim \cite{cite20} in a Quasi-Poisson distribution. Time series analysis is often adopted to quantitatively evaluate the short-term correlation between environmental factors such as temperature, air pollution, and health indicators. Independent variables fitting were carried out on the air pollutants (PM$_{2.5}$, PM$_{10}$, SO$_2$, CO, NO$_2$, and O$_3$) and daily median air temperature and humidity for the DLNM model. Estimating that the daily confirmed incidence of the population is a small probability event, and there may be over discretization phenomenon, uses Quasi-Poisson regression as the connection function of the model. More specifically, to assess the possibility of seasonality and air pollutants' impact on COVID-19, the participants include Sydney, Melbourne, Brisbane, Seoul, Gwangju, Gyeonggi-do, Trieste (only used in humidity), Brescia, Milan, Rome. Consequently, the hierarchy analysis approach was adopted, where air pollutants and seasonality data are incorporated into the DLNM model.

We carried out this study in DLNM model and also assumed that the effects of PM$_{10}$, PM$_{2.5}$, and SO$_2$ are linear with R$_0$ (fun = "lin") and the effects of O$_3$, CO, and NO$_2$ are null up to their respective minimum ranges and linear with R$_0$ and applying three low thresholds parameterization (fun = "thr"). The argvar list is applied to generate the matrix for the space of the predictor while modeling the relationship with temperature and humidity through a natural cubic spline with three degrees of freedom (fun = "ns", chosen by default). Conceptually, the modified DLNM model is shown as Equation (1), (2), and (3):

$\log[E(Y_t)]= \alpha + cb (Median Temp_t, lag, argvar = ("ns", df = 3), arglag = ("ns", knots = logknots(lag,1))) + cb (Median Humi_t,lag,argvar = ("ns",df = 3), arglag = ("ns", knots = lgoknots(lag,1)))$ (1)

where Y$_t$ represents the value of R$_0$ on the day of t; $\alpha$ represents the intercept; cb represents the cross basis; lag represents the most prolonged lag period; Temp$_t$ represents the daily median temperature on day t and Humi$_t$ represents the daily median humidity on day t. The knots for the spline of lags are placed at equally-spaced values in the log scale of lags, using the function logknots.

$\log[E(Y_t)]= \alpha + cb (Median AP_t, lag, argvar  = ("lin"))$ (2)

where AP$_t$ represents the concentration of pollutants on the day of t. To PM$_{10}$, PM$_{2.5}$ and SO${_2}$, they are incorporated into the model depending on the linear effect. The argvar list in which the function generates a basis matrix, including a linear un-transformed variable and applying one parameter for each lag up to 7 days.

$\log[E(Y_t)]= \alpha + cb (Median AP_t, lag, argvar = ("thr"))$ (3)

Supposing O$_3$, CO, and NO$_2$ were depending on the bilinear threshold effect, which applied one parameter for each lag up to 7 days.

In this investigation there are several sources for limitation.  The main limitation is the lack of pollutant data in Trieste. Another major source of uncertainty is in the method used to estimate lag. Due to the calculable R$_0$ sequence is short, we can not accurately estimate lag values. All of the above models were fitted using the Nonlinear effect of distribution lag, as implemented in the R statistical software package.



%%       \citet{<label>} ==> Jones et al. [21]
%%       \citep{<label>} ==> [21]
%%



\section{Results}
The first set of analyses examined the trend of R$_0$ in 10 cities. South Korea, Italy, and Australia have issued strict control measures on February 27th, March 21st, and March 31st, 2020, to control the spread of infection. Due to the characteristic of \cite{cite5}, we adopted February 27th, March 21st, and March 31st as the expiry dates while calculating R$_0$ of three countries. From January 31st to February 27th, 2020, 56 confirmed cases were reported in Seoul, 62 in Gyeonggi-do, and 9 in Gwangju. From March 4th to 21st, 2020, there were 4672 confirmed cases in Milan, 5028 in Brescia, 893 in Rome, and 270 in Trieste. From March 1st to March 31st, 2020, 2032 confirmed cases were reported in Sydney, 917 in Melbourne, and 743 in Brisbane. To estimate R$_0$, repeated-measures EpiEstim \cite{cite20} were used. Figure 1, 2, and 3 shows an overview of R$_0$ in ten cities.

Air pollutant index levels are known to vary considerably across industrial development. Table 1 shows the average air index of nine cities in Australia, South Korea, and Italy. It is apparent from this table that air pollutant indexes in each city in Italy (except CO) and South Korea towards a significantly higher than those in Australia.
% Table generated by Excel2LaTeX from sheet 'Sheet3'

The significant correlation differences between Australia, South Korea, and Italy are highlighted in Figure 4 and Figure A.1-A.8. To perspective three cities in Australia, the temperature has a significant positive correlation with PM$_{10}$ and PM$_{2.5}$, humidity has a significant negative correlation with PM$_{10}$, and PM$_{10}$ has a significant positive correlation with PM$_{2.5}$, and SO$_2$, O$_3$, CO, NO$_2$ do not correlate. There was a significant positive correlation between PM$_{10}$ and PM$_{2.5}$ in two Italian cities, but no correlation among other air indexes. From three cities of data in South Korea, the temperature has a significant positive correlation with CO and NO$_2$, PM$_{10}$ has a significant positive correlation with PM$_{2.5}$, SO$_2$, CO, and NO$_2$, PM$_{2.5}$ has a significant positive correlation with SO$_2$, CO, and NO$_2$, SO$_2$ has a significant positive correlation with NO$_2$, CO has a significant positive correlation with NO$_2$, and a significant negative correlation with O$_3$, NO$_2$ has a significant negative correlation with O$_3$. It is apparent from this table that PM$_{10}$ and PM$_{2.5}$ in Australia and Italy cities are a significantly positive correlation, in which except for Milan, the temperature has a significant positive correlation with PM$_{10}$ and PM$_{2.5}$. The most surprising aspect of the data is in the six pollutants in Seoul, where any two pollutants have a significant correlation.

In the seasonal analysis, to analyze the impact of temperature and humidity on R$_0$, we chose Brisbane (maximum temperature), Gyeonggi-do (minimum temperature), Gyeonggi-do (maximum humidity),  Trieste (minimum humidity) and adopted relative risk (RR) as the evaluation criteria. Figure 5 is quite revealing in several ways. First, the distribution of relative risk (RR) varies with temperature and humidity. Secondly, ${27.6}^{\circ}\mathrm{C}$ and ${8.5}^{\circ}\mathrm{C}$ are the optimum temperature (lowest RR) for Brisbane and Gyeonggi-do, respectively. Above these temperatures, the higher the RR. Below these temperatures, the greater the RR. 59$\%$ and 28$\%$ are the optimum humidity (lowest RR) of Gyeonggi-do and Trieste, respectively. Above these humidities, the higher the humidity, the higher the RR. Below these humidities, the lower the humidity, the higher the RR. From Table 2 below, we can see that the short-term impact of air temperature and humidity on R$_0$ has apparent lag and persistence, but there was no evidence that R$_0$ is simply linear with adjustments in temperature and humidity.

% Table generated by Excel2LaTeX from sheet 'Sheet4'
The correlation coefficients of R$_0$ and air pollutants were examined in nine cities, and 5$\%$ was chosen as the significance level. Strong evidence of primary pollutant was found when PM$_{10}$ passing the correlation coefficient examine. Further statistical tests revealed that CO and NO$_2$ are the lowest indicators passing the significance test. To evaluate the interaction between different pollutants, and O$_3$, PM$_{2.5}$, and SO$_2$ are incorporated into the DLNM model as perturbation factors. Air pollutants effect was manifested by Relative Risk (RR), and Excess morbidity Rate (ER), ER= (RR-1) * 100$\%$ denotes the percentage increase of R$_0$ caused by 14ug/m$^{3}$ of PM$_{10}$. The themes identified in these responses are summarized in Table 3.

% Table generated by Excel2LaTeX from sheet 'Sheet2'

\section{Discussion}
In this paper, we have used a distributed lag nonlinear model for estimating the time-varying effects of air temperature, humidity, and pollution on daily R$_0$. The model combines information across multiple cities to compare the variation of seasonal relative effect rate estimates. We discovered that the influence of air temperature and humidity on short-term R$_0$ has apparent lag and persistence. Temperature and humidity patterns varied by geographic region, the fluctuation of R$_0$ with temperature and humidity is not linear. Our results suggest that it has a particular influence on the growth of R$_0$ in a range of lag. Furthermore, the most prominent finding to emerge from the analysis is that PM$_{10}$ as the primary pollutant affecting the growth of R$_0$, which the rise of PM$_{10}$ concentration will cause an increase of ER, and O$_3$, PM$_{2.5}$, and SO$_2$ as perturbation factors also have a particular impact on ER.

The modification of effects of virus by season and air pollution has been explored previously in several single-city studies. Figure 2 illustrates those of Hu et al. \cite{cite21} who also investigated the short-term effect and lag effect of air environment on disease risk. Styer et al. \cite{cite24} analyzed data from Cook County, Illinois, and Salt Lake County, Utah, and found (for Cook County) that the effect of PM$_{10}$ was higher in the spring and fall. The estimation of the short-term impact of air pollution on a single city's daily confirmed case is hindered by the high variability inherent in the impact estimates. Estimating the impact of seasonal changes is an additional challenge as it involves further stratification of the data. Liao et al. \cite{cite23} investigated the short-term impact of air pollutants on the number of ILs in Yichang City, China, by applying the distributed lag nonlinear model, and indicated that there was no statistical significance in the impact of pollutants in winter. Since the early outbreak of COVID-19 is mainly concentrated in the northern hemisphere, there are few data available to estimate the seasonal effect, so the variability of this estimation increases, making it difficult to identify any significant seasonal pattern. Instead, we chose Australia in autumn, South Korea, and Italy in the spring season, to describe the natural framework of regional and national trends. Furthermore, the concentration of air pollutants in Italy and South Korea is much higher than that in Australia, which indicates a distinct contrast. Based on the statistical results in Table 2, Seoul is the city with the highest concentration of PM$_{10}$ and PM$_{2.5}$, and one of the cities with a more severe epidemic. However, we find in Table 4 shows that the confidence interval is varying under the impact of air pollutants, and summarizes the impact of nine city pollutants on ER. These estimates are based on the results of the DLNM model when the PM10 concentration drives by 14ug/m$^{3}$. Our results suggest that the ER value of Seoul was the highest (1027.93, 95$\%$ confidence interval of RR was 3.40-311740.5) under the union of PM$_{10}$ and PM$_{2.5}$. The single most striking observation to emerge from Italy and South Korea data comparison was that cities with severe pollution have the most significant impact on the ER. There are similarities between the attitudes expressed by the higher concentration of pollutants cause of the higher incidence rate in this study and those described by Moolgavkar\cite{cite22}.

It should be noted that this study has investigated only reviewed the epidemic data of Australia, South Korea, and Italy because each country immediately offered some control measures at the same time as the epidemic outbreak. However, we require to calculate the basic reproduction rate (R$_0$) that before the control effect. Unfortunately, we can not determine a relatively long R$_0$ sequence from other countries' data. However, these issues could be solved if we consider the anti-seasonality of the northern and southern hemispheres; therefore, we chose Italy and South Korea, which has a relatively long R$_0$ sequence, and different from countries in the southern hemisphere in terms of season and pollution level. The wide confidence interval of air pollutants is interesting, but not surprising. The research gap in this study toward the impact of environmental pollution on COVID-19 involves time-series data. Due to each pollution index tends to have a frequent variation with time and the amount of samples is inadequate, multiple collinearity is easy to associate between them. In the process of regression, the higher the correlation between independent variables, the more severe the multicollinearity, so the more significant the variance of the estimated value of the regression coefficient, the wider the confidence interval \cite{draper1998applied}. We adopted the DLNM model to forecast and ensures that the types of independent variables in the future are consistent, and the relationship between independent variables in the next modeling still has the characteristics of original data. Consequently, although multiple collinearity variables are included in the regression model, and solid prediction results can also be obtained.
\section{Conclusions}
This paper has analyzed the relationship between environment and R$_0$ by using the DLNM model based on Quasi-Poisson regression. The findings of this study suggest that R$_0$ is affected by seasonal variation (temperature and humidity) and has apparent lag and persistence, and seasonal factors have an apparent decorating effect on short-term R$_0$, and PM$_{10}$ has the most significant effect on PM$_{10}$ daily. These findings in this report are subject to at least three limitations. First, R$_0$ sequences are relatively short, which is challenging to evaluate lag days. Second, some cities need to be evaluated, which their atmospheric data are missing. Third, there is not sufficient sample data to assess the impact of four seasons on COVID-19. A further study could assess the long-term effects of non-homogeneous populations and establish more complex models.
\section{Conflicts of interest}
The authors declared no competing interest.
\section{Acknowledgments}
% \newpage
\appendix

\section{Supplementary data}
\begin{figure}[h!]
  \centering
  \includegraphics[width=.6\columnwidth]{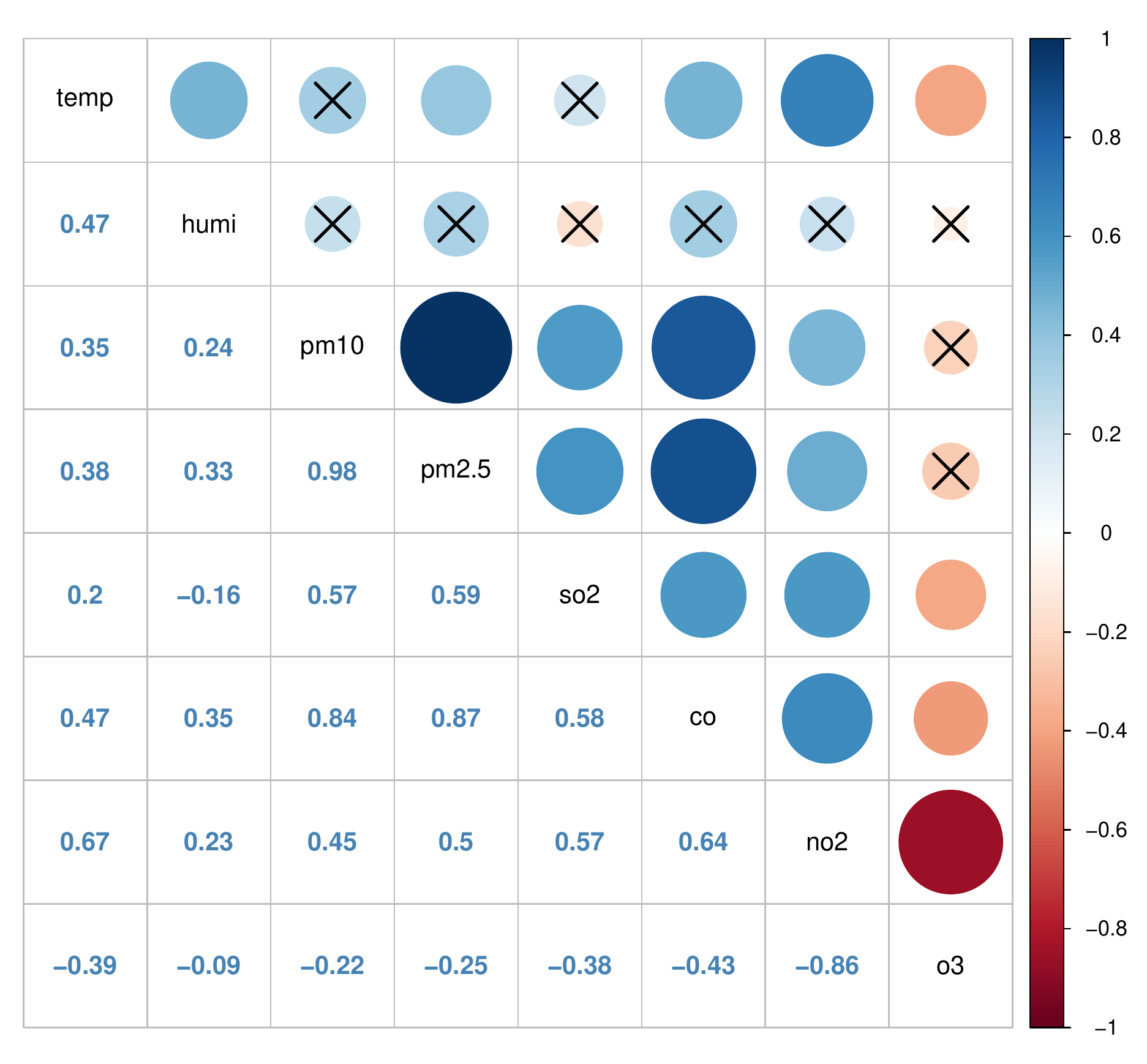}
  \caption{Spearman correlation analysis of temperature, humidity and air pollution (Gyeonggi-do)}
  \label{fig:Gyeonggi-do-res}
\end{figure}

\begin{figure}
  \centering
  \includegraphics[width=.6\columnwidth]{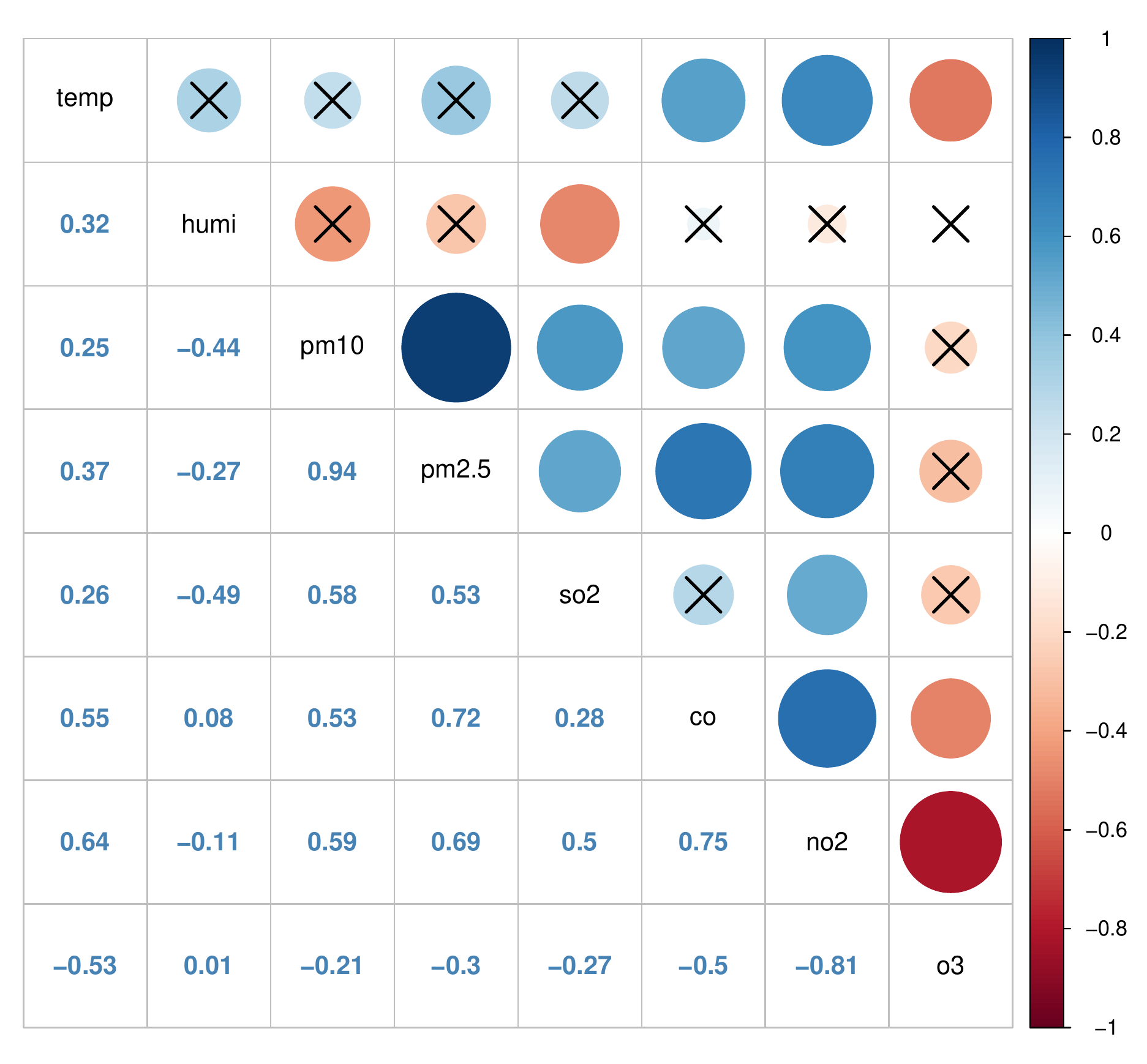}
  \caption{Spearman correlation analysis of temperature, humidity and air pollution (Gwangju)}
  \label{fig:Gwangju-res}
\end{figure}

\begin{figure}
  \centering
  \includegraphics[width=.6\columnwidth]{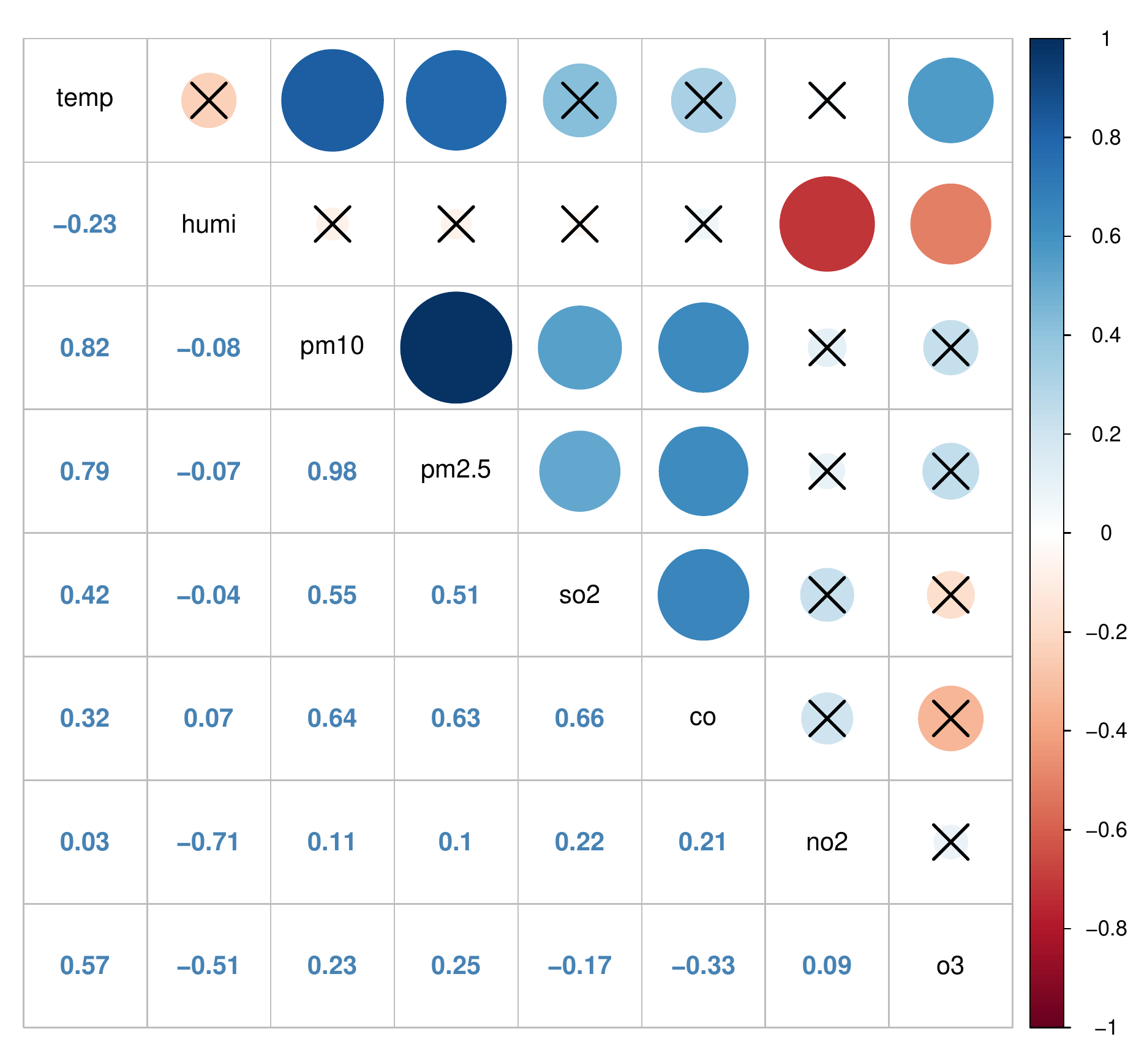}
  \caption{Spearman correlation analysis of temperature, humidity and air pollution (Brescia)}
  \label{fig:Brescia-res}
\end{figure}

\begin{figure}
  \centering
  \includegraphics[width=.6\columnwidth]{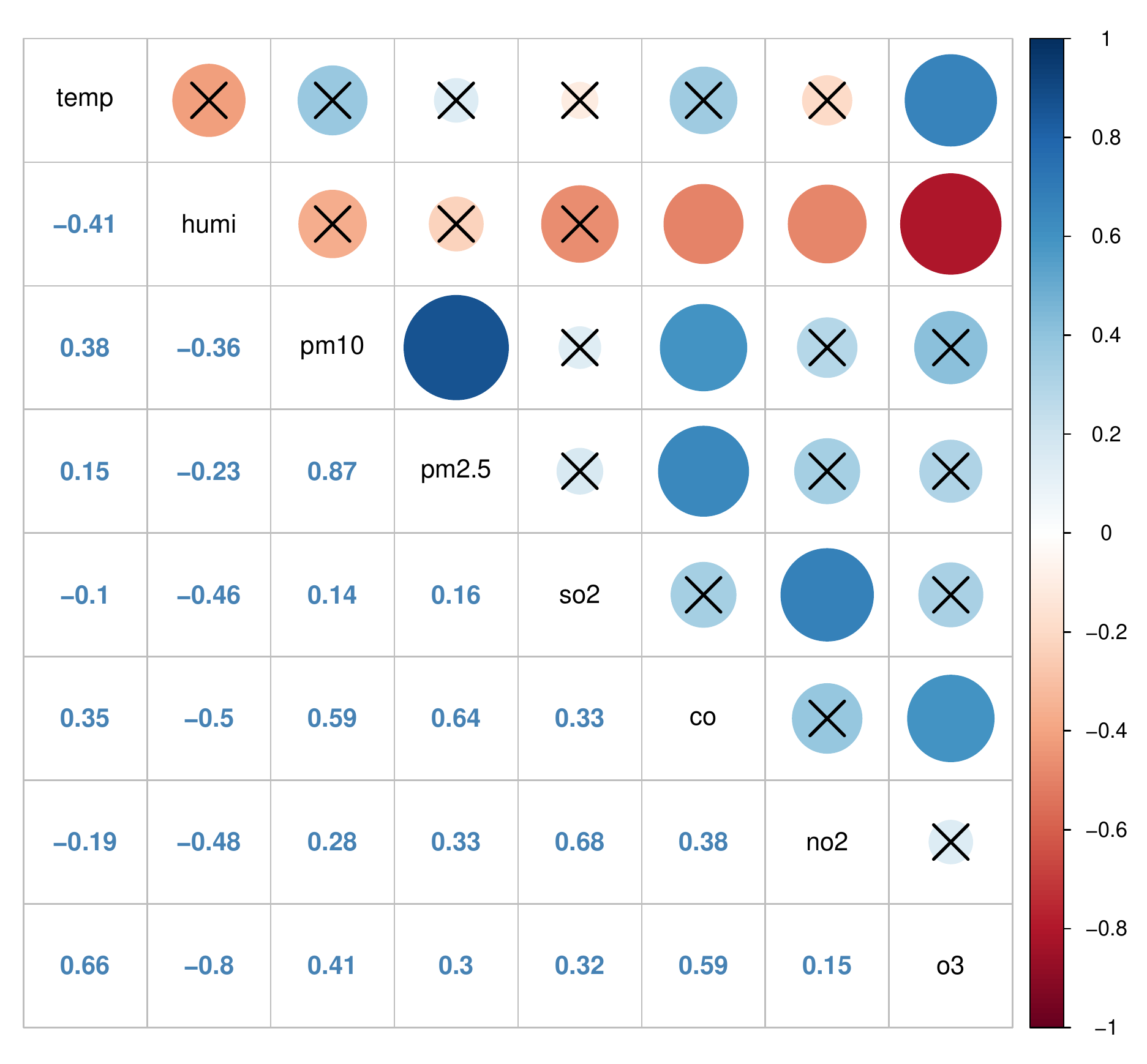}
  \caption{Spearman correlation analysis of temperature, humidity and air pollution (Milan)}
  \label{fig:Milan-res}
\end{figure}

\begin{figure}
  \centering
  \includegraphics[width=.6\columnwidth]{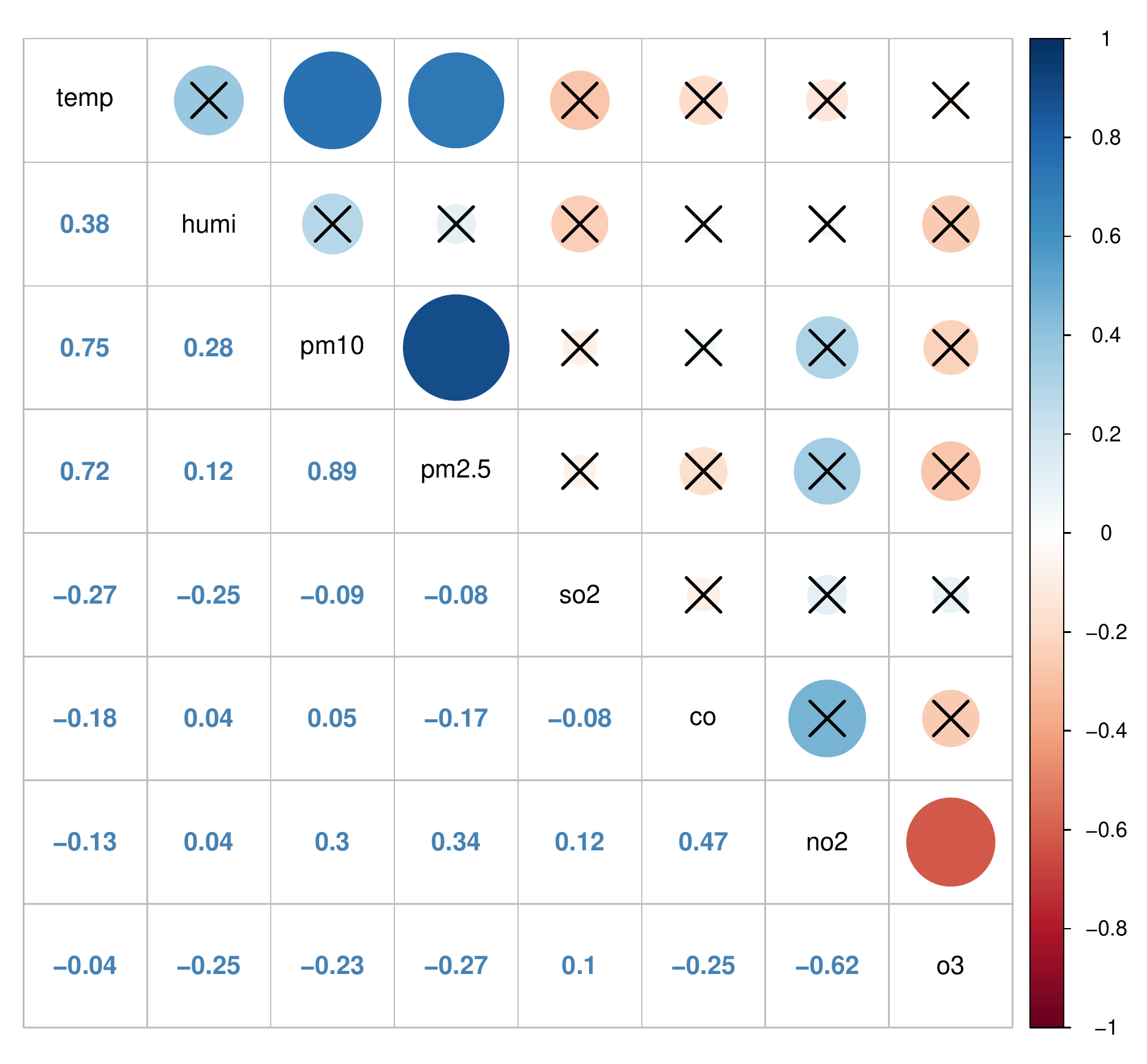}
  \caption{Spearman correlation analysis of temperature, humidity and air pollution (Rome)}
  \label{fig:Rome-res}
\end{figure}

\begin{figure}
  \centering
  \includegraphics[width=.6\columnwidth]{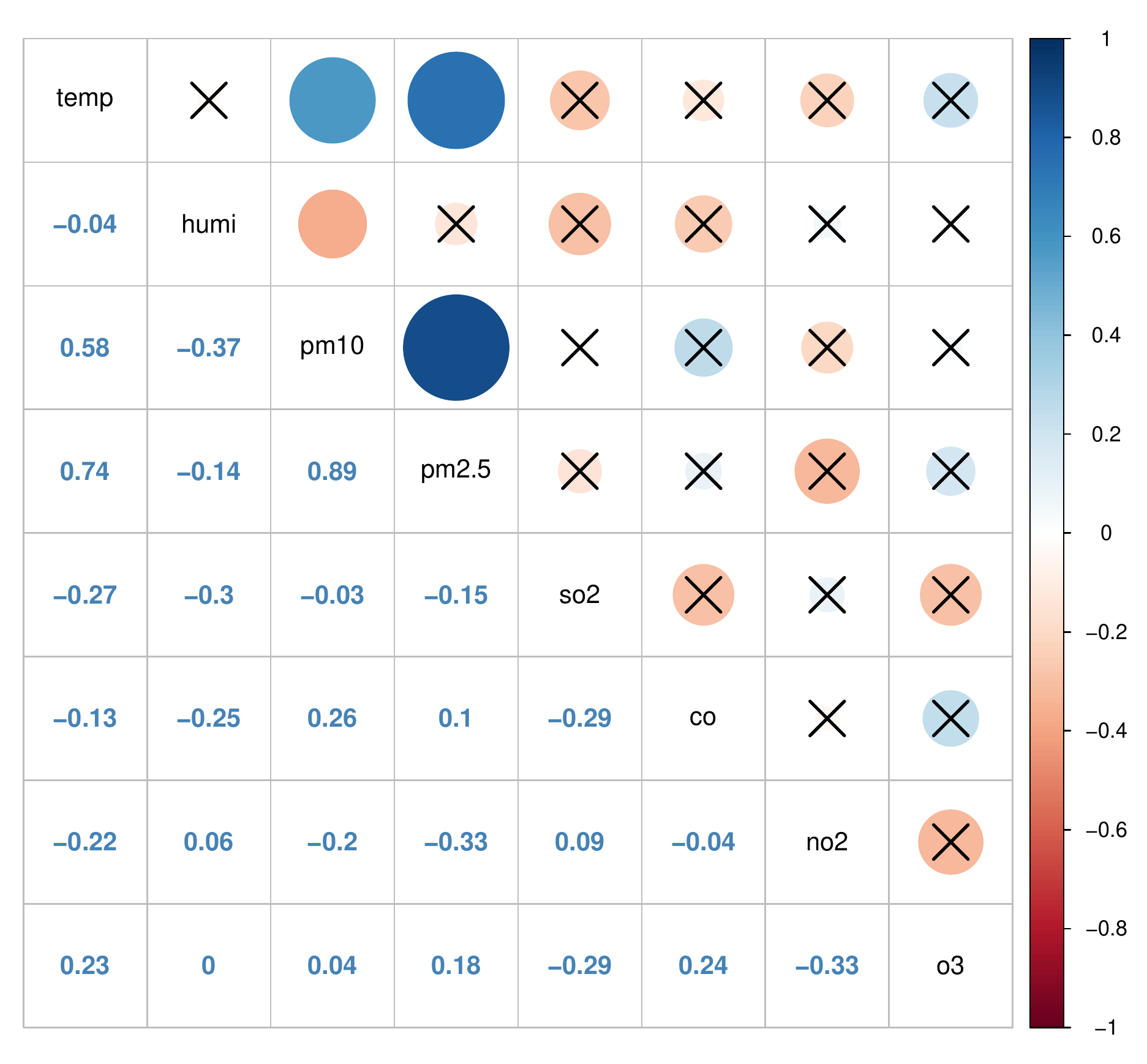}
  \caption{Spearman correlation analysis of temperature, humidity and air pollution (Brisbane)}
  \label{fig:Brisbane-res}
\end{figure}

\begin{figure}
  \centering
  \includegraphics[width=.6\columnwidth]{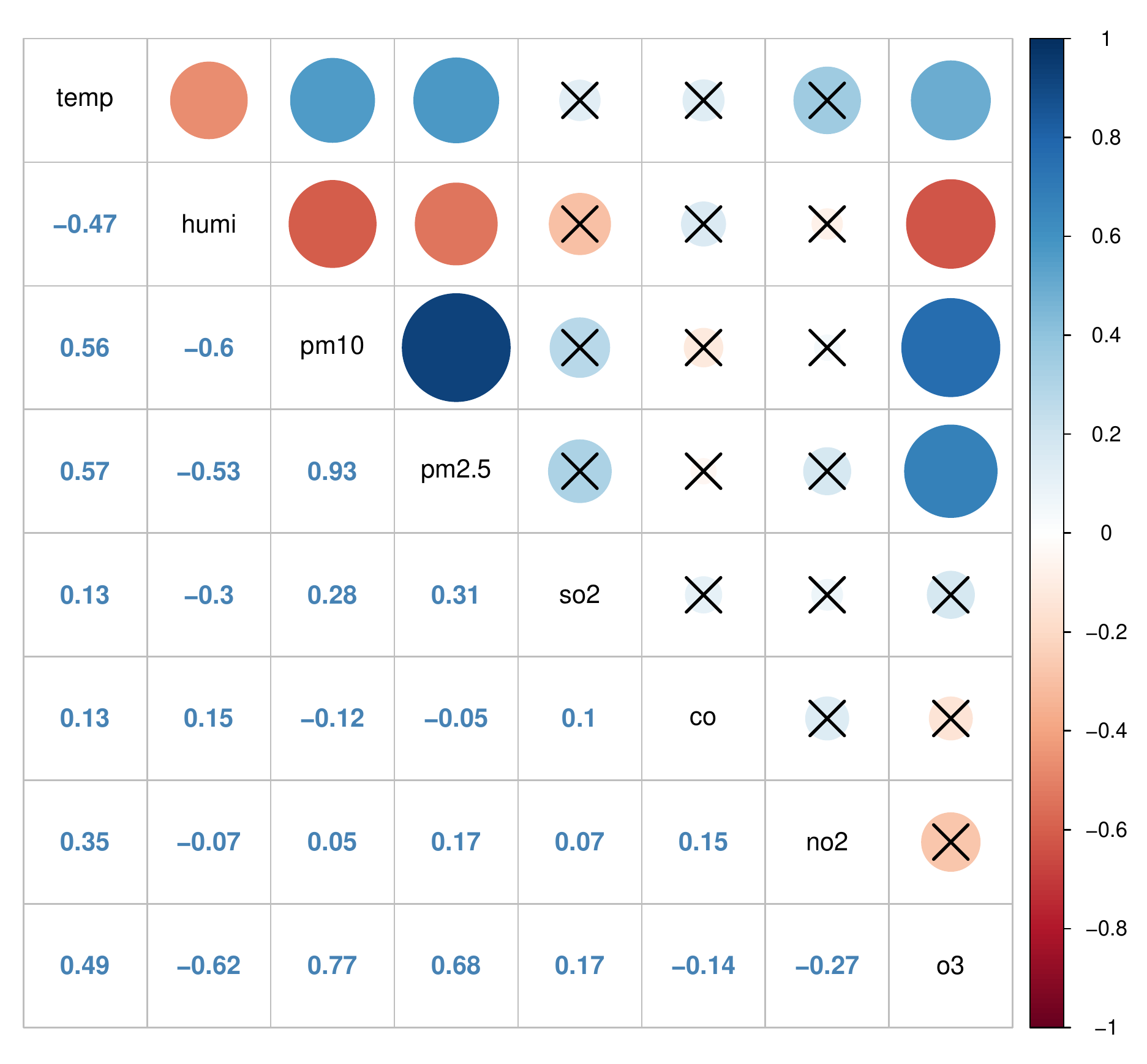}
  \caption{Spearman correlation analysis of temperature, humidity and air pollution (Sydney)}
  \label{fig:sydney-res}
\end{figure}

\begin{figure}
  \centering
  \includegraphics[width=.6\columnwidth]{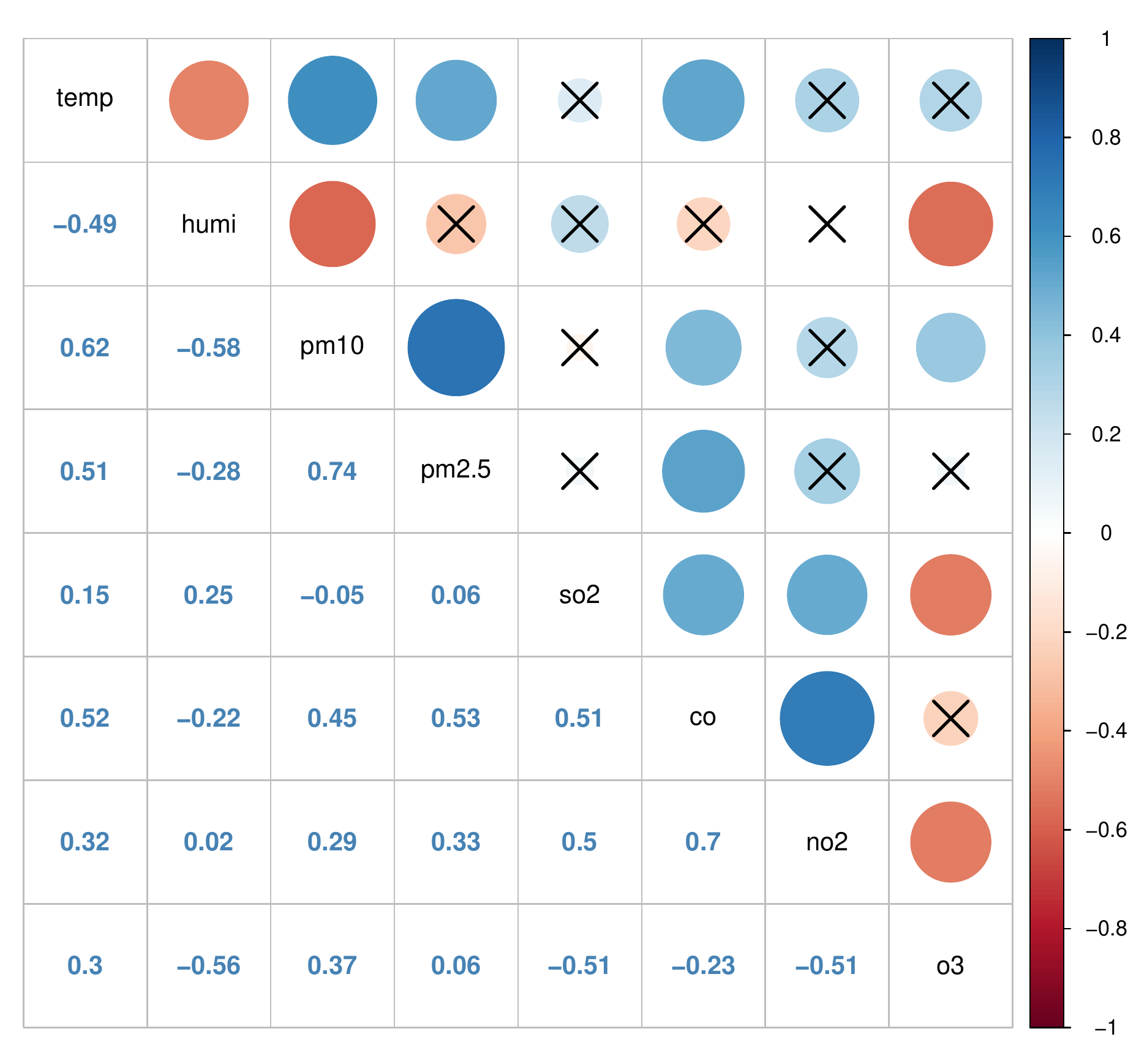}
  \caption{Spearman correlation analysis of temperature, humidity and air pollution (Melbourne)}
  \label{fig:Melbourne-res}
\end{figure}

\clearpage
\section*{}

\bibliographystyle{elsarticle-harv}
\bibliography{citation}

\newpage
\counterwithout{figure}{section}
\counterwithout{table}{section}
\setcounter{figure}{0}
\begin{figure}
  \centering
  \includegraphics[width=.89\columnwidth]{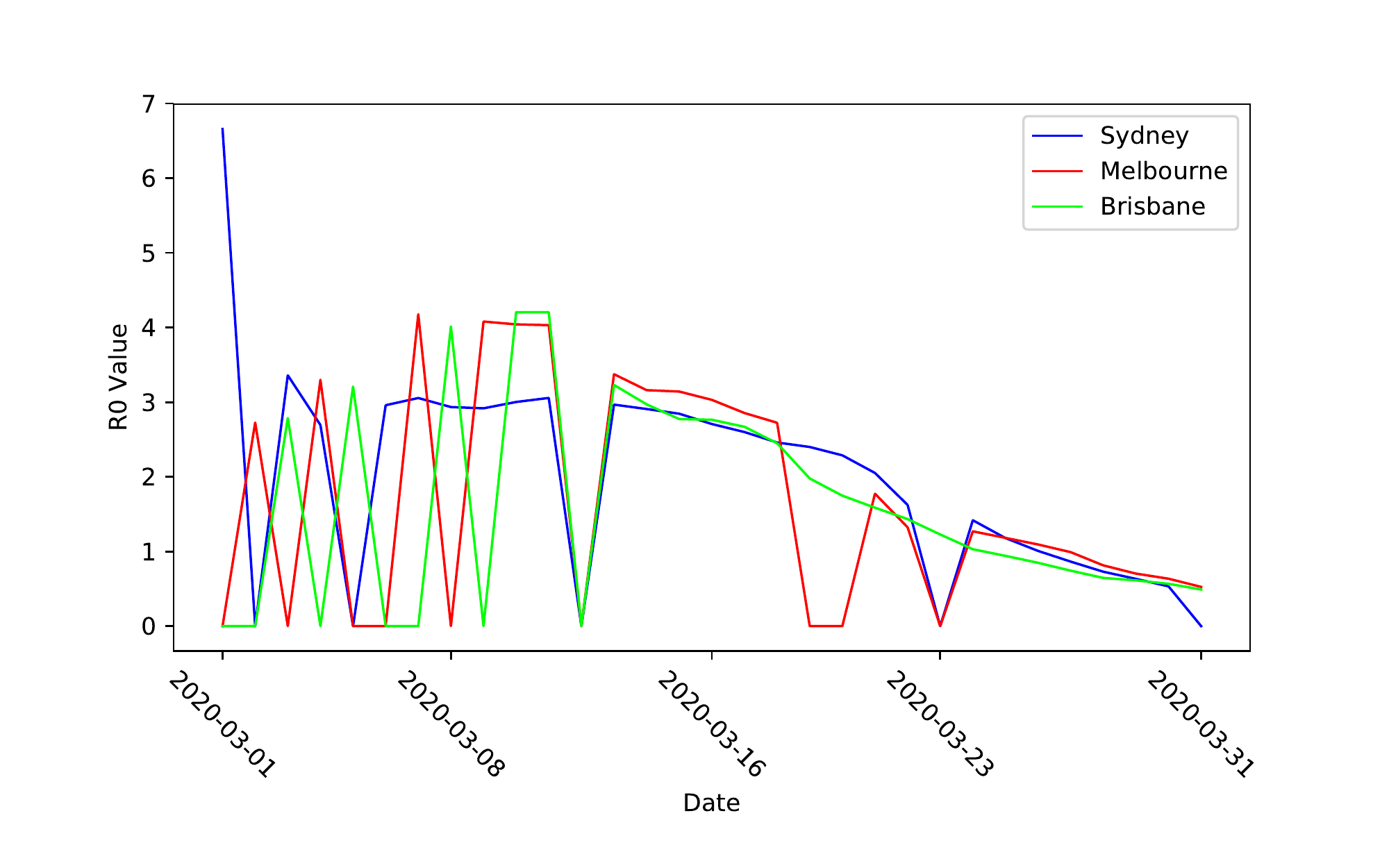}
  \caption{The variation of R$_0$ with date in Australia.}
  \label{fig:aus-r0}
\end{figure}

\begin{figure}
  \centering
  \includegraphics[width=.89\columnwidth]{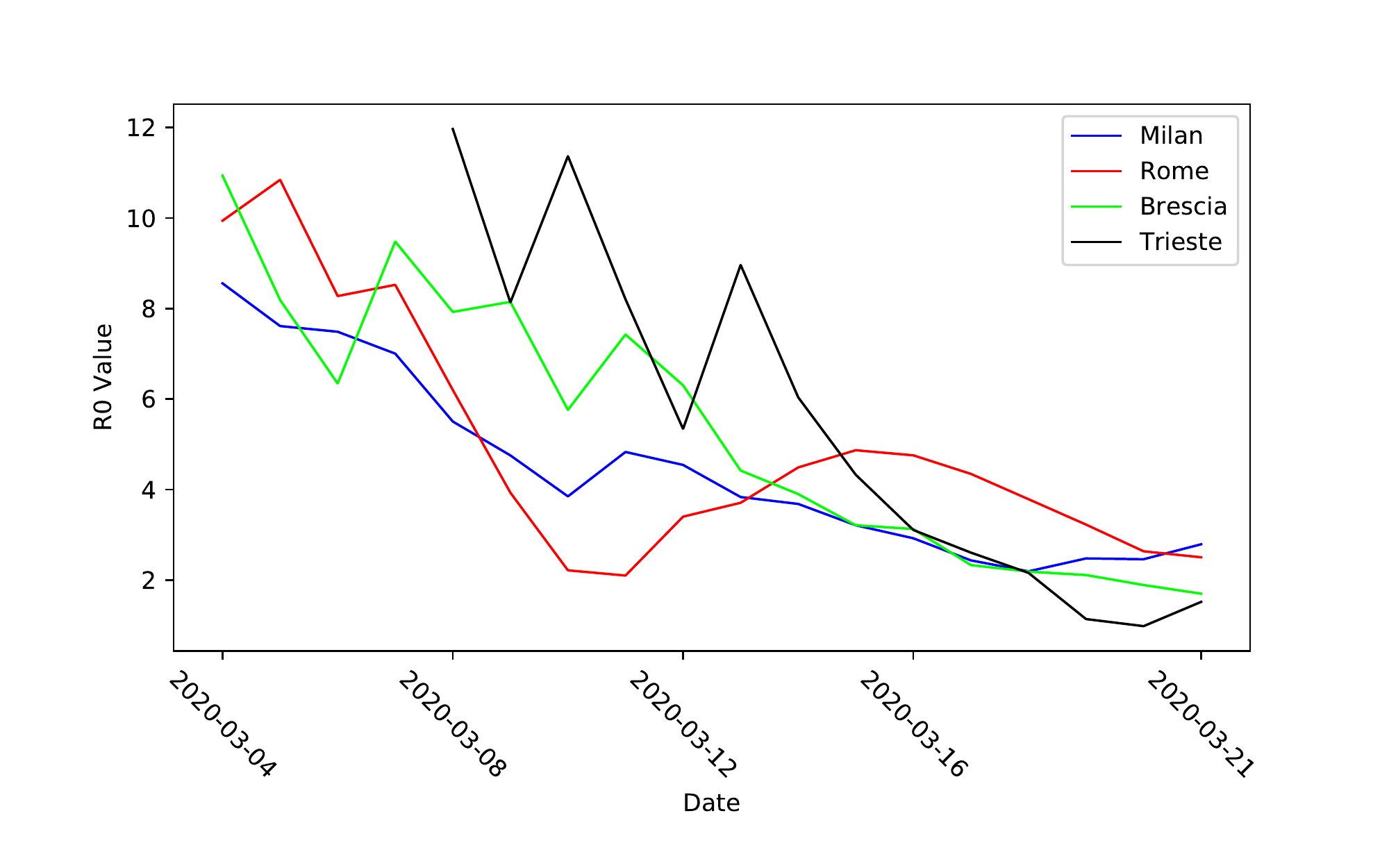}
  \caption{The variation of R$_0$ with date in Italy.}
  \label{fig:ita-r0}
\end{figure}

\begin{figure}
  \centering
  \includegraphics[width=.89\columnwidth]{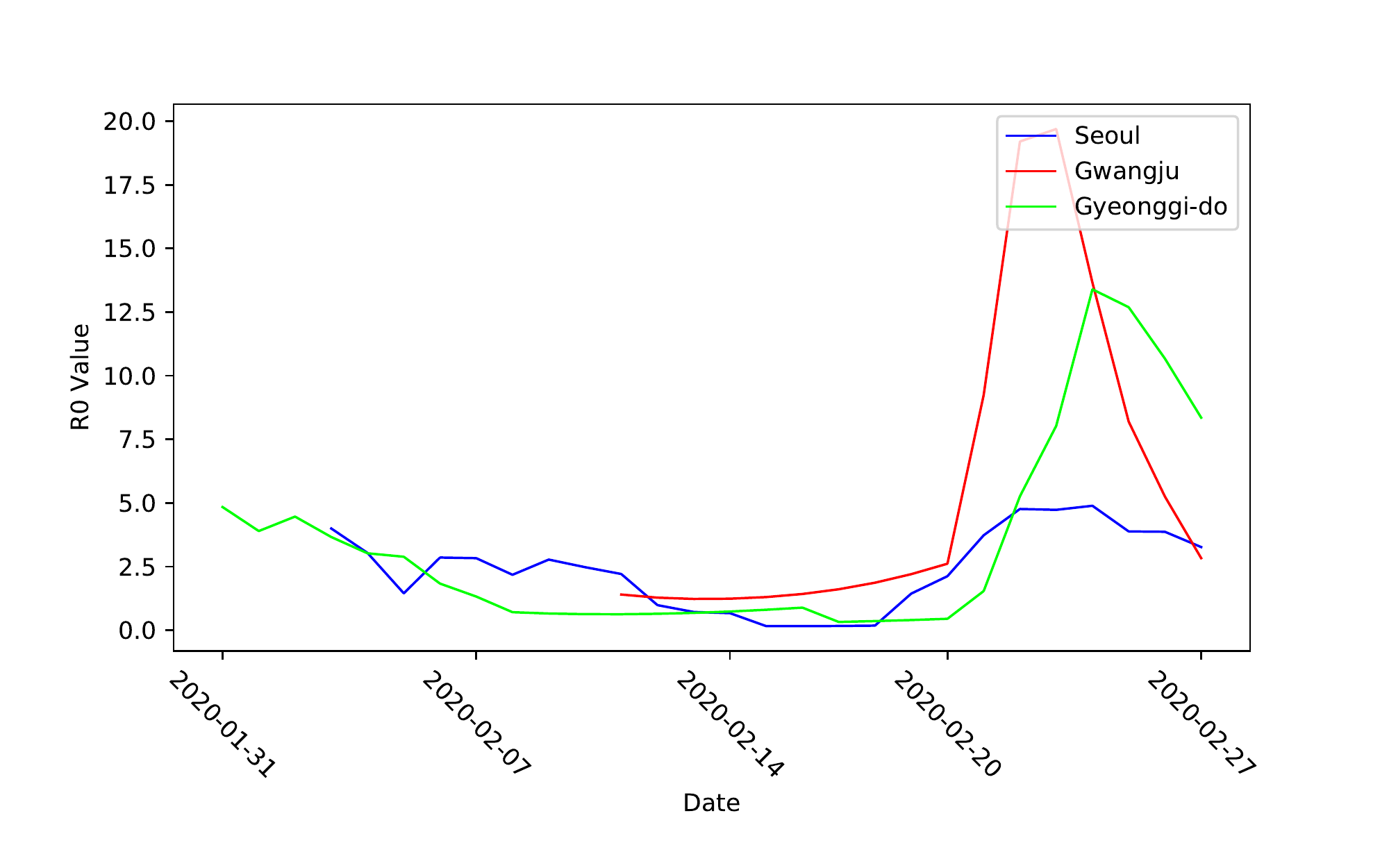}
  \caption{The variation of R$_0$ with date in South Korea.}
  \label{fig:kor-r0}
\end{figure}

\begin{figure}
  \centering
  \includegraphics[width=.6\columnwidth]{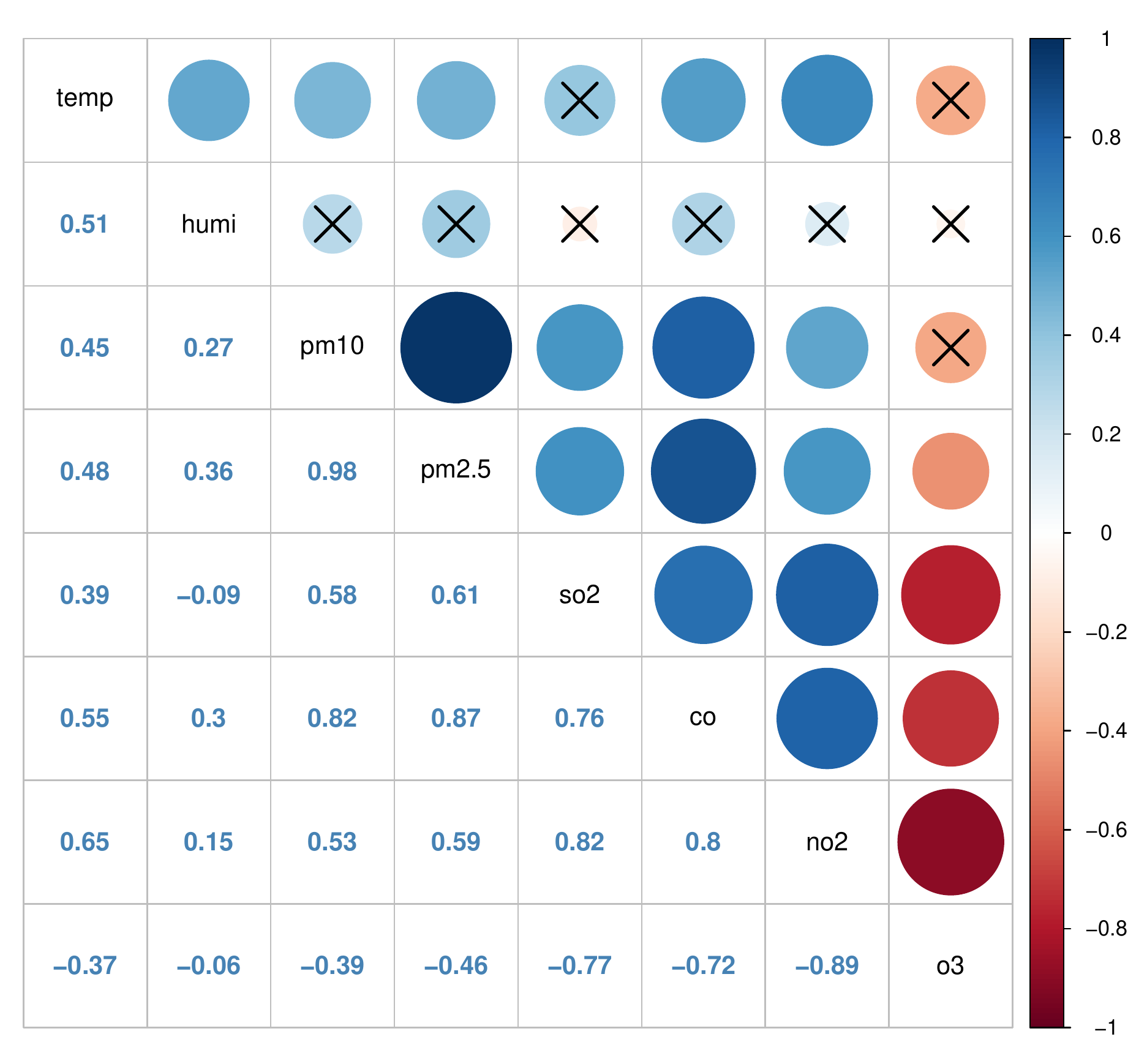}
  \caption{Spearman correlation analysis of temperature, humidity and air pollution (Seoul)
  The upper triangle panel uses circles of different colors and sizes to represent the correlation; Blue represents the positive correlation; Red represents the negative correlation; The darker the color, the larger the circle represents the higher the correlation, with 5$\%$ as the significance level, and the result declining to pass the statistical correlation test is X; the lower triangle panel uses the value of correlation coefficient to represent the correlation direction and size;}
  \label{fig:Seoul-res}
\end{figure}

\begin{figure}
  \centering
  \includegraphics[width=.95\columnwidth]{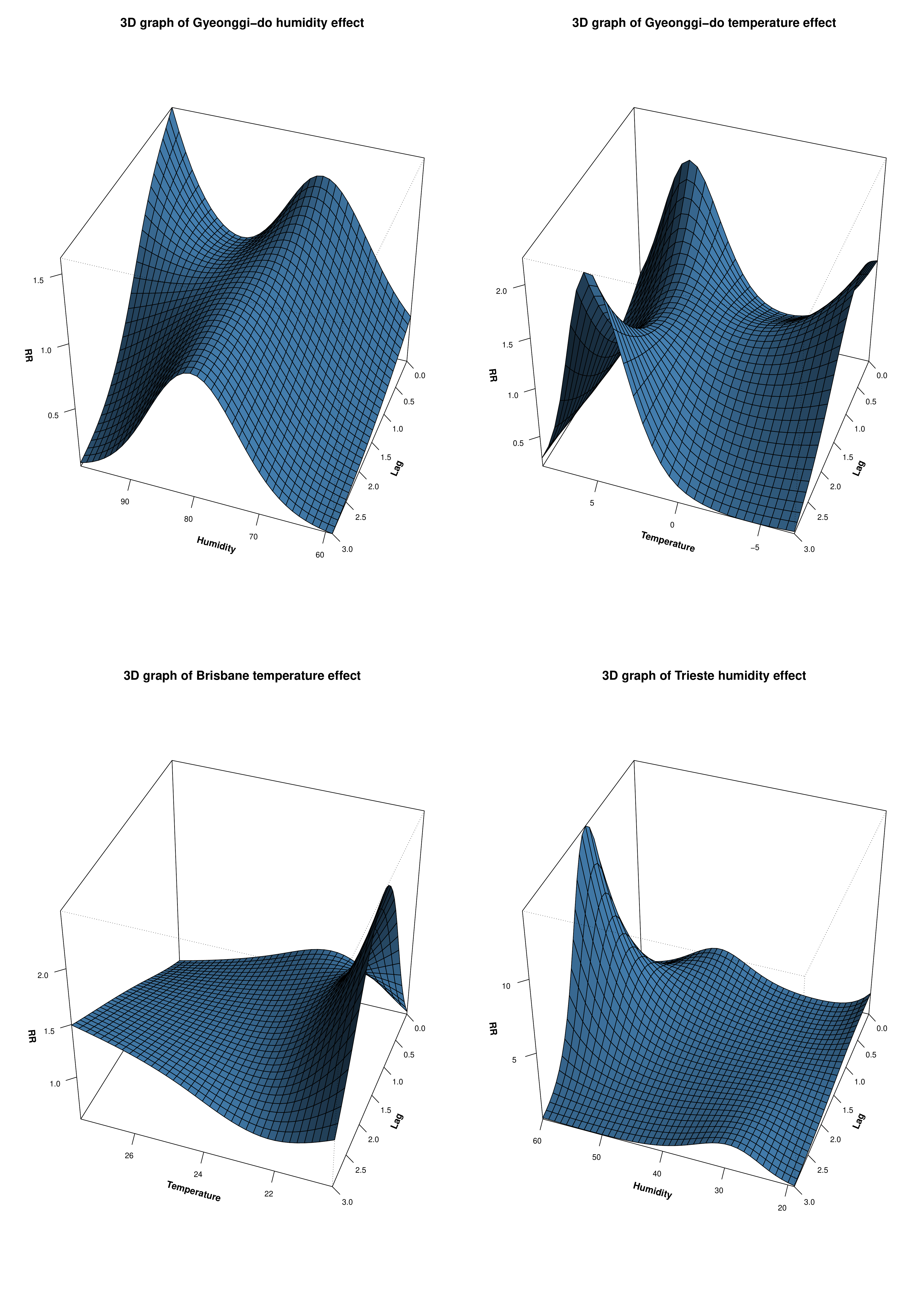}
  \caption{The variation of relative risk (RR) with humidity and lag time (lag) 3D-graph.}
  \label{fig:3d}
\end{figure}

\clearpage
\begin{table}[htbp]
  \footnotesize
  \centering
  \caption{Average of temperature, humidity and pollutants in nine cities.}
  \begin{adjustbox}{width=\textwidth}
    \begin{tabular}{cccccccccc}
    \toprule
    Index & Sydney & Melbourne & Brisbane & Milan & Rome & Brescia & Seoul & Gwangju & Gyeonggi-do \\
    \midrule
    Temperature & 20.54 & 17.98 & 23.48 & 10.69 & 12.63 & 10.08 & 2.29 & 6.91 & 2.17 \\
    Humidity & 77.13 & 64.56 & 70.62 & 67.53 & 71.11 & 60.04 & 56.58 & 72.96 & 80.46 \\
    PM$_{10}$ & 14.97 & 15.19 & 11.9 & 28.5 & 21.44 & 25.89 & 35.4 & 27.47 & 35.25 \\
    PM$_{2.5}$ & 17.26 & 19.09 & 19.26 & 73.67 & 53.3 & 70.94 & 79.52 & 58.24 & 73.75 \\
    SO$_{2}$ & 1.74 & 0.35 & 1.42 & 2.71 & 0.38 & 1.52 & 5.08 & 3.72 & 5.68 \\
    CO   & 2.02 & 0.95 & 1.99 & 0.79 & 0.47 & 0.6  & 6.44 & 5.6  & 6.74 \\
    NO$_{2}$ & 6.09 & 6.03 & 3.64 & 31.6 & 11.45 & 23.76 & 29.09 & 15.84 & 24.72 \\
    O$_{3}$   & 13.9 & 12.42 & 6.91 & 33.67 & 18.65 & 32.13 & 14.45 & 20.43 & 18.24 \\
    \bottomrule
    \end{tabular}
  \end{adjustbox}
  \label{tab:avgdata}%
\end{table}

\begin{table}[htbp]
  \centering
  \footnotesize
  \caption{Association between temperature and humidity (daily increase of temperature and humidity levels in different lag days) and RR.}
  \begin{adjustbox}{width=\textwidth}
    \begin{tabular}{ccccc}
    \toprule
    Multi-day lag & Max Temperature & Min Temperature & Max Humidity & Min Humidity \\
    (Cumulative effects) & RR(95\%CI) & RR(95\%CI) & RR(95\%CI) & RR(95\%CI) \\
    \midrule
    0-0  & 1.03(0.76,1.42) & 0.95(0.89,1.01) & 1.03(0.97,1.10) & 1.01(1.00,1.02) \\
    0-1  & 2.09(1.59,2.74) & 1.93(1.83,2.04) & 2.04(1.94,2.15) & 2.01(2.00,2.02) \\
    0-2  & 3.10(2.43,3.97) & 2.90(2.76,3.05) & 3.04(2.90,3.20) & 3.01(2.99,3.02) \\
    0-3  & 4.04(3.15,5.20) & 3.82(3.62,4.03) & 4.05(3.83,4.28) & 4.00(3.98,4.02) \\
    \bottomrule
    \end{tabular}
  \end{adjustbox}
  \label{tab:table2}%
\end{table}

\begin{table}[htbp]
  \centering
  % \footnotesize
  \setlength\tabcolsep{1pt}
  \caption{In the model of multiple pollutants, air pollutants causes the ER.}
  \begin{adjustbox}{width=\textwidth}
    \begin{tabular}{ccccccc}
    \toprule
    Air pollutants & Sydney &      & Melbourne &      & Brisbane &  \\
         & ER   & 95\%CI & ER   & 95\%CI & ER   & 95\%CI \\
    \midrule
    PM$_{10}$ & 0.11 & 1.03-1.20 & 0.31 & 1.05-1.63 & -0.25 & 0.45-1.27 \\
    PM$_{10}$+O$_3$ & 0.35 & 1.13-1.61 & 0.13 & 0.80-1.58 & 0.16 & 0.69-1.94 \\
    PM$_{10}$+PM$_{2.5}$ & -0.36 & 0.32-1.29 & 0.26 & 0.94-1.68 & 0.26 & 0.08-18.82 \\
    PM$_{10}$+SO$_2$ & 0.03 & 0.95-1.13 & 0.05 & 0.76-1.44 & 0.13 & 0.56-2.31 \\
    PM$_{10}$+PM$_{2.5}$+O$_3$+SO$_2$ & -0.82 & 0.03-1.00 & -0.07 & 0.59-1.49 & -0.78 & 0.01-5.90 \\
    \midrule
         &      &      &      &      &      &  \\
    \midrule
    Air pollutants & Milan &      & Rome &      & Brescia &  \\
         & ER   & 95\%CI & ER   & 95\%CI & ER   & 95\%CI \\
    \midrule
    PM$_{10}$ & 0.87 & 0.87-4.01 & 0.14 & 0.77-1.69 & 5.93 & 3.22-14.92 \\
    PM$_{10}$+O$_3$ & -0.77 & 0.04-1.20 & 0.43 & 1-2.05 & 2.25 & 1.78-5.93 \\
    PM$_{10}$+PM$_{2.5}$ & 178.84 & 0.96-337444 & 9.71 & 1.00-144.87 & -0.99 & 0-0.14 \\
    PM$_{10}$+SO$_2$ & 0.1  & 0.64-1.88 & 0.29 & 0.60-2.78 & 5.23 & 3.05-12.69 \\
    PM$_{10}$+PM$_{2.5}$+O$_3$+SO$_2$ & 0    & 0.01-123.72 & 2.72 & 0.32-43.54 & -0.99 & 0-1.16 \\
    \midrule
         &      &      &      &      &      &  \\
    \midrule
    Air pollutants & Seoul &      & Gwangju &      & Gyeonggi-do &  \\
         & ER   & 95\%CI & ER   & 95\%CI & ER   & 95\%CI \\
    \midrule
    PM$_{10}$ & 11.3 & 3.23-46.86 & -0.83 & 0.01-3.34 & 48.94 & 3.90-639.57 \\
    PM$_{10}$+O$_3$ & 0.53 & 0.33-7.22 & 0.91 & 0.10-35.76 & 39.69 & 1.20-1381.88 \\
    PM$_{10}$+PM$_{2.5}$ & 1027.93 & 3.40-311740.5 & -0.92 & 0-2.82 & -0.33 & 0-5277.72 \\
    PM$_{10}$+SO$_2$ & 2.75 & 0.79-17.77 & -0.63 & 0-31.82 & 5.1  & 0.27-136.69 \\
    PM$_{10}$+PM$_{2.5}$+O$_3$+SO$_2$  & 24347.04 & 5.66-104745200 & 113431.1 & 188.53-68249940 & -1   & 0-19.42 \\
    \bottomrule
    \end{tabular}
  \end{adjustbox}
  \label{tab:table3}%
\end{table}

\clearpage

\section*{HIGHLIGHTS}
\begin{itemize}

	\item{Air temperature and humidity have lag and persistence on short-term R$_0$.}
	\item{Seasonal factors have an apparent decorating effect on R$_0$.}
    \item{PM$_{10}$ is the primary pollutant that affects the excess morbidity rate.}
    \item{O$_{3}$, PM$_{2.5}$, and SO$_{2}$ as perturbation factors have cumulative effect on ER.}

\end{itemize}




% \end{thebibliography}
\end{document}